\documentclass[aps,twocolumn,preprintnumbers,amsmath,amssymb,floatfix]{revtex4}
\usepackage{epsfig}
\usepackage{graphicx}
\usepackage{bm}
\usepackage{amssymb}
\usepackage{dcolumn}
\usepackage{bm}
\begin{document}
\draft

\title
{Fluorescence detection of single lithium atoms in an optical lattice using Doppler-cooling beams}
\author{Hyok Sang Han, Hyun Gyung Lee, Seokchan Yoon, and D. Cho\footnote{e-mail address:{\tt cho@korea.ac.kr}}}
\affiliation{Department of Physics, Korea University, Seoul 02841, Korea}

\date{\today}

\begin{abstract}
We demonstrate {\it in situ} fluorescence detection of $^7$Li atoms in a 1D optical lattice with single atom precision.
Even though illuminated lithium atoms tend to boil out, when the lattice is deep, molasses beams without extra cooling retain the atoms while producing sufficient fluorescent photons for detection.
When the depth of the potential well at an antinode is 2.4 mK, an atom remains trapped for 30 s while scattering probe photons at the rate of $1.7 \times 10^5$ s$^{-1}$.
We propose a simple model that describes the dependence of the lifetime of an atom on well depth.
When the number of trapped atoms is reduced, a clear stepwise change is observed in integrated fluorescence, indicating the detection of a single atom.
At a photon-collecting efficiency of only 1.3\% owing to small numerical aperture, the presence or absence of an atom is determined within 300 ms with an error of less than $5 \times 10^{-4}$.
\end{abstract}

\maketitle

\section{INTRODUCTION}
As both internal and motional states of trapped atoms are controlled more and more precisely for their quantum manipulation, it has also become very important to observe those atoms {\it in situ} and individually. Efforts to image fluorescence from single trapped atoms started with those in a magneto-optical trap (MOT) \cite{Kimble}.
It was extended to single atoms in a 1D optical lattice with a site-specific resolution using a diffraction-limited imaging system and careful offline analysis \cite{Meschede2009}.
These efforts culminated when the individual sites of a 2D optical lattice were imaged using objective lenses with high numerical aperture (NA) \cite{Greiner2009,Kuhr2010}.
The 2D version is known as a quantum-gas microscope, and it was developed primarily to prepare samples for and read out results from quantum simulation of interacting particles.
Initially, these experiments were performed using either $^{133}$Cs \cite{Kimble,Meschede2009} or $^{87}$Rb \cite{Greiner2009,Kuhr2010} atoms because molasses beams can be used to simultaneously image and cool heavy alkali-metal atoms.
In recent years, 2D imaging techniques have been extended to fermionic atoms such as $^{6}$Li \cite{Greiner2015} and $^{40}$K \cite{Zwierlein2015,Kuhr2015}, which are better proxies for strongly-interacting electrons.
However, light atoms tend to boil out before scattering sufficient photons for imaging because of their large recoil energy and poor polarization gradient cooling.
To overcome this difficulty, Raman sideband cooling \cite{Greiner2015,Zwierlein2015} and electromagnetically-induced-transparency (EIT) cooling \cite{Kuhr2015} have been employed. This complicates the apparatus and imaging process.
In addition, an exposure time of longer than 1 s is required because Raman cooling and EIT cooling rely on putting atoms in low-lying dark states. The energy-lowering stimulated processes are interlaced with brief optical-pumping stages, during which photons are harvested.

In the present work, using only Doppler-cooling beams, we demonstrate {\it in situ} imaging of single $^7$Li atoms in a 1D optical lattice with single atom precision.
Lattice depth $U_0$ turns out to be a critical parameter; above $U_0 = 1.5$ mK, there is an abrupt increase in the number of photons scattered by an atom before it escapes the lattice. A simple model of evaporation followed by Doppler cooling explains this phenomenon. Although the nearest sites are not resolved in our detection because of small NA of 0.22, our approach can be combined with either a large-NA system or spectroscopic identification of individual sites \cite{MeschedeQR} to facilitate quantum gas microscopy of light atoms. In our measurement at $U_0$ = 2.4 mK, the presence or absence of an atom can be determined with 99.95\% probability using a 300-ms exposure time, despite the low photon-collecting efficiency.

\section{APPARATUS}
A double MOT fed by a Zeeman slower is used to load lithium atoms to an optical lattice \cite{magic-pol}.
The 1D lattice is formed in an octagonal glass chamber by focusing and retro-reflecting a Gaussian beam. See Fig. 1. The wavelength $\lambda_L$ is 1064 nm and the $e^{-2}$ intensity radius at the focus is 14 $\mu$m. Mode matching of the reflected beam is optimized by maximizing the power coupled back to the optical fiber that delivers the lattice beam.
When incident power is 1.3 W, the depth $U_0$ at an antinode is 1 mK or 830$E_R$, where $E_R = h^2/2m\lambda_L^2$.  A home-built ytterbium-doped fiber laser provides the single-frequency lattice beam.
MOT beams with a radius of 1.6 mm are used as imaging beams.
The fluorescence from lattice atoms is collected by an objective lens with NA of 0.22 and refocused to an electron-multiplying charge-coupled device (EMCCD) with unit magnification.
NA of 0.22 corresponds to a photon-collecting efficiency of 1.3\% and the EMCCD has a quantum efficiency of 90\% at 671 nm.
With further reduction by 0.9 owing to scattering and diffraction losses, one out of 100 fluorescent photons are detected \cite{MOT-imaging}.

\begin{figure}[h] 
	\includegraphics[scale=0.4]{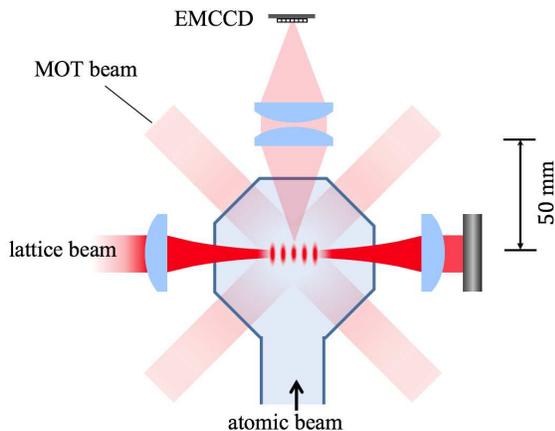}
	\caption{(Color online) Fluorescence-imaging system for the 1D optical lattice formed in an octagonal glass chamber. EMCCD: electron-multiplying charge-coupled device.} 
\end{figure}

\section{EXPERIMENT}
Our aim is to detect the fluorescence from lattice-bound atoms with single atom precision.
We collect data from a region of interest (ROI), which consists of 3 by 3 pixels of the EMCCD.
Each pixel measures $16 \times 16$ $\mu\rm{m}^2$, and the ROI corresponds to 100 sites at the center of the lattice.
In the first part of the experiment, we attempt to determine the conditions that allow {\it{in situ}} imaging of atoms using Doppler-cooling beams. In the second part, we reduce the number of atoms to observe stepwise change in integrated fluorescence.

Typically, we load a thousand atoms to the lattice using the MOT with low-power beams of 150 $\mu$W in each direction for both trapping and repumping. An anti-Helmholtz coil is turned off and the MOT beams optimized for imaging are illuminated.
For $^7$Li, the scalar polarizabilities of the $2S_{1/2}$ and $2P_{3/2}$ states at $\lambda_L$ = 1064 nm are -270 and -167 in atomic units, respectively \cite{JYKim, Safranova}. The $2P_{3/2}$ state has a negative polarizability owing to its coupling to the $3S_{1/2}$ and $3D$ states, and it is a trappable state.
Nevertheless, the $|2S_{1/2}, F=2 \,\rangle$ $\rightarrow$ $|2P_{3/2}, F=3 \,\rangle$ transition suffers both frequency shift and inhomogeneous broadening; the lattice beam causes a  blue shift of 8 MHz in the $D2$ transition when $U_0$ is 1 mK.
Detuning of the MOT trap beam is adjusted for a given $U_0$ to maximize the number of photons $N_c$ scattered by an atom before it escapes from the lattice. The repump beam is stabilized to the $|2S_{1/2}, F=1 \,\rangle$ $\rightarrow$ $|2P_{3/2}, F=2 \,\rangle$ transition with fixed detuning.
Illumination of the near-resonant beams results in sites with either one or no atoms owing to photoassociative losses.
We use approximately 50 atoms trapped at the central 100 sites for the fluorescence detection.

\begin{figure}[h] \centering
	\includegraphics[scale=0.4]{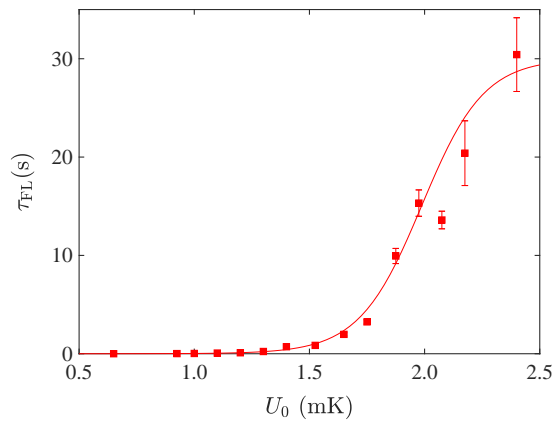}
	\caption{(Color online) Decay time constant $\tau_{FL}$ of the fluorescence signal from the trapped lithium atoms vs. well depth $U_0$ of the lattice. $\tau_{FL}$ increases abruptly for $U_0$ larger than 1.5 mK and it saturates at the vacuum-limited lifetime of 30 s. The fitting curve is obtained using a model of alternating ejection and Doppler cooling processes.}
\end{figure}

The decay time constant $\tau_{FL}$ of the fluorescence signal versus $U_0$ is plotted in Fig. 2.
For this measurement, we use 25 $\mu$W in each direction for the trap and repump beams.
$\tau_{FL}$ is related to the number of scattered photons as $N_c = R \tau_{FL}$, where $R$ is the photon scattering rate.
Because of the inhomogeneous broadening, $R$ depends on $U_0$ and it is on the order of $10^5$ s$^{-1}$.
When $U_0$ is below 1.4 mK, Li atoms escape the lattice almost instantaneously because of large recoil energy and lack of polarization gradient cooling, as noted earlier.
However, above $U_0 = 1.5$ mK, $\tau_{FL}$ increases steeply and at $U_0 = 2.4$ mK it reaches 30 s, which is the vacuum-limited trap lifetime $\tau_{VC}$.
This behavior may be explained by assuming that the independent processes of boil out and Doppler cooling alternate while atoms are imaged.
Suppose that the lattice atoms are Doppler cooled to temperature $T_D$ and those with energy higher than $U_0$ are ejected. The ejected fraction $p$ is $(1+\theta+\theta^2/2)e^{-\theta}$ with $\theta = U_0/k_B T_D$ when the 3D harmonic potential well is isotropic \cite{Ketterele1995}.
In the proposed model, this ejection occurs instantaneously resulting in a truncated Maxwell-Boltzmann distribution. This is followed by another round of Doppler cooling, which lasts for  $\tau_{MB}$ to reestablish the distribution at $T_D$. This model leads to the following relation:
\begin{equation}
\frac{1}{\tau_{FL}} = \frac{1}{\tau_{VC}} -\frac{\ln (1-p)}{\tau_{MB}}.
\end{equation}
The fitting curve shown in Fig. 2 is for $T_D = 120$ $\mu$K, $\tau_{MB} = 300$ $\mu$s, and $\tau_{VC} = 30$ s.
We note that the Doppler cooling limit for lithium is 140 $\mu$K.
In the experiment on 2D imaging of $^6$Li in Ref. \cite{Greiner2015}, $U_0$ at the antinode is estimated to be 0.8 mK from the quoted trap frequencies, and the Raman sideband cooling was indispensable for imaging.

In order to demonstrate the single-atom precision, we load only one or two atoms to the central part of the lattice by reducing the loading time and waiting until the ROI signal decreases beyond a threshold. We maintain $U_0$ at 2.4 mK and the imaging power at 25 $\mu$W. We continue to use the ROI of 3 by 3 pixels, but we take data only when the center pixel is the brightest so that the imaged atoms are within $\pm$10 sites from the minimum spot. The Rayleigh range of the lattice beam is 580 $\mu$m, and these sites can be considered identical.

\begin{figure}[t] \centering
	\includegraphics[scale=0.4]{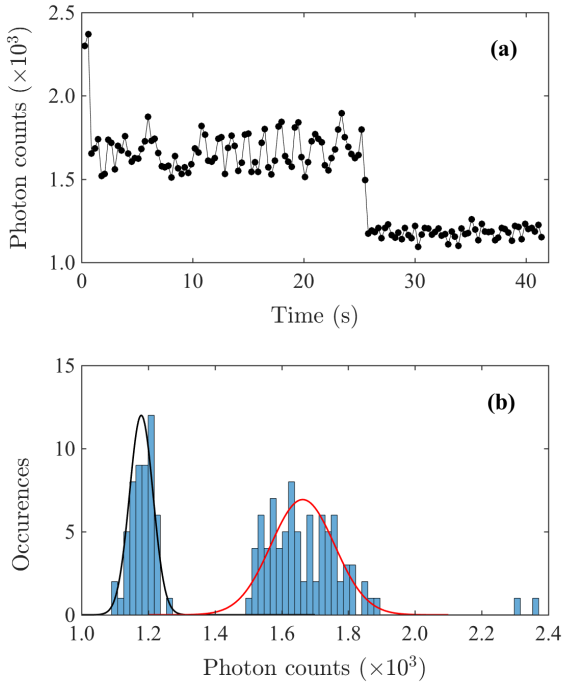}
	\caption{(Color online) Detection of single atoms in the lattice. (a) Time evolution of the fluorescence signal as the lattice loses one atom at a time. The exposure time for each data point is 300 ms. (b) Histogram of the photon counts during the exposure time in (a). Gaussian fits for the zero (black) and one atom (red) cases are included.}
\end{figure}

Figure 3 (a) shows the evolution of the ROI signal as the lattice loses one atom at a time, starting with two atoms.
The exposure time for each data point is 300 ms. In this measurement, the single atom stays trapped for 25 s.
The count rate of the fluorescence from a single atom is 1.7 kHz and that from the scattered imaging beam is 3.8 kHz. We note that the fluorescence count rate corresponds to the photon scattering rate $R= 1.7 \times 10^5$ s$^{-1}$ and it is only 6\% of that from a single atom in a MOT with the same imaging beam intensities \cite{MOT-imaging}. We attribute this reduction to the inhomogeneous broadening of the $D2$ transition in the lattice.
The histogram of the time-series data is shown in Fig. 3 (b).
Using the Gaussian fits, we obtain an average and standard deviation of 1180 and 35, respectively, for the zero-atom case, and 1660 and 95, respectively, for the one-atom case.
As was already apparent in Fig. 3 (a), the fluorescence signal exhibits considerable noise, which was not observed in the MOT fluorescence \cite{MOT-imaging}.
We conclude that this noise is caused by fluctuation in lattice-beam parameters such as power and mode matching.
Even with the noise, when we use a proper threshold, which is 1340 counts in this case, both the sensitivity of deciding correctly the presence of a single atom and the specificity of deciding correctly the absence of an atom are larger than 99.95\% for an exposure time of 300 ms. With the large NA of the quantum gas microscopes, the exposure time can be reduced by more than an order of magnitude.

\section{Conclusion}
In this work, we address the problems in imaging or detecting lithium atoms in an optical lattice. When the potential well is sufficiently deep, simple Doppler cooling enables us to obtain a fluorescence signal with single-atom precision while the atoms remain trapped. We plan to combine this capability with the site-specificity obtained by using an applied magnetic field gradient to achieve site-resolved {\it in situ} imaging for our quantum manipulation experiment on single lithium atoms.

\section*{ACKNOWLEDGMENTS}
This work was supported by a grant to the Atomic Interferometer Research Laboratory for National Defense funded by the Defense Acquisition Program Administration and Agency for Defense Development in Korea.

\newpage



\begin{thebibliography}{99}
\bibitem{Kimble}
Z. Hu and H. J. Kimble, Opt. Lett. {\bf 19}, 1888 (1994).
\bibitem{Meschede2009}
M. Karski, L.F\"{o}rster, J. M. Choi, W. Alt, A. Widera, and D. Meschede, Phys. Rev. Lett. {\bf 102}, 053001 (2009).
\bibitem{Greiner2009}
W. S. Bakr, J. I. Gillen, A. Peng, S. F\"{o}lling, and M. Greiner, Nature (London) {\bf 462}, 74 (2009).
\bibitem{Kuhr2010}
J. F. Sherson, C. Weitenberg, M. Endres, M. Cheneau, I. Bloch, and S. Kuhr, Nature (London) {\bf  467}, 68 (2010).
\bibitem{Greiner2015}
M. F. Parsons, F. Huber, A. Mazurenko, C. S. Chiu, W. Setiawan, K. Wooley-Brown, S. Blatt, and M. Greiner, Phys. Rev. Lett. {\bf 114}, 213002 (2015).
\bibitem{Zwierlein2015}
L. W. Cheuk, M. A. Nichols, M. Okan, T. Gersdorf, V. V. Ramasesh, W. S. Bakr, T. Lompe, and M. W. Zwierlein, Phys. Rev. Lett. {\bf 114}, 193001 (2015).
\bibitem{Kuhr2015}
E. Haller, J. Hudson, A. Kelly, D. A. Cotta, B. Peaudecerf, G. D. Bruce, and S. Kuhr, Nature Phys. {\bf 11}, 738 (2015).
\bibitem{MeschedeQR}
D. Schrader, I. Dotsenko, M. Khudaverdyan, Y. Miroshnychenko, A. Rauschenbeutel, and D. Meschede, Phys. Rev. Lett. {\bf 93}, 150501 (2004).
\bibitem{magic-pol}
H. Kim, H. S. Han, and D. Cho, Phys. Rev. Lett. {\bf 111}, 243004 (2013).
\bibitem{MOT-imaging}
H. S. Han, S. Yoon, and D. Cho, J. Korean Phys. Soc. {\bf 66}, 1675 (2015).
\bibitem{JYKim}
J. Y. Kim, J. S. Lee, J. H. Han, and D. Cho, J. Korean Phys. Soc. {\bf 42}, 483 (2003).
\bibitem{Safranova}
M. S. Safronova, U. I. Safronova, and C. W. Clark, Phys. Rev. A {\bf 86}, 042505 (2012).
\bibitem{Ketterele1995}
K. B. Davis, M. -O. Mewes, W. Ketterle, Appl. Phys. B {\bf 60}, 155 (1995).

\end{thebibliography}
\end{document}